\documentclass[	
						superscriptaddress,%
						twocolumn,%
						a4paper,
						showkeys,
					]{revtex4}

\usepackage[english]{babel}
\usepackage{upgreek}
\usepackage{textcomp} 				
\usepackage[dvips]{graphicx}		

\newcommand{\etal}{\textit{et al.}}
\newcommand{\eps}{\varepsilon}

\begin{document}

\title{High piezoelectric coefficient of single domain Mn-doped NBT-6\%BT single crystals}
\date{\today}
\author{Mael Guennou}
\author{Maxim Savinov}
\author{Jan Drahokoupil}
\affiliation{Institute of Physics, Academy of Sciences of the Czech Republic, Na Slovance 2 18221, Prague 8, Czech Republic}
\author{Haosu Luo}
\affiliation{Shanghai Institute of Ceramics, Chinese Academy of Sciences, 215 Chengbei Rd., Jiading 201800, Shanghai, China}
\author{Jirka Hlinka}
\affiliation{Institute of Physics, Academy of Sciences of the Czech Republic, Na Slovance 2 18221, Prague 8, Czech Republic}

\begin{abstract}
We report a study of properties of Mn-doped NBT-6\%BT single crystals. We show that tetragonal single domain states can be stabilized by poling along a $[001]$ direction. For carefully prepared crystals, the piezoelectric coefficient $d_{33}$ can reach 570~pC/N. When poled along non-polar directions, the crystals exhibit ferroelectric domain structures consistent with tetragonal micron-sized domains, as revealed by optical observation and Raman spectroscopy. The multidomain crystals have lower $d_{33}$ values, 225 and 130~pC/N for $[011]$ and $[111]$-oriented crystals respectively. This trend is commented on from a domain-engineering perspective.
\end{abstract}


\maketitle


In the search for lead-free piezoelectric materials, the (Na$_{1/2}$Bi$_{1/2}$)$_{1-x}$Ba$_x$TiO$_3$ (NBT-$x$BT) system has attracted a lot of attention both for its good performances in terms of piezoelectric constant and electromechanical coupling factor and its relaxor behavior \cite{Roedel2009,Panda2009,Shvartsman2012}. The composition-temperature phase diagram of the NBT-$x$BT system shows a quasi-vertical region that separates a rhombohedral-like side and a tetragonal side around $x\approx 6$\% \cite{Takenaka1991,Jo2011}. Although this is strongly reminiscent of the morphotropic phase boundary (MPB) in lead-based system (PZT, PMN-PT, PZN-PT etc.), for which the best electromechanical properties to date are known, the underlying mechanisms are very different. No monoclinc bridging phase has been identified so far in the NBT-BT system; instead, the MPB is described in terms of coexistence of rhombohedral and tetragonal phases, and the piezoelectric enhancement is believed to be related to polarization extension rather than polarization rotation \cite{Ge2012}. In addition, the delicate crystal structures in the MPB is strongly affected by the electric field during the poling process \cite{Jo2011,Ma2012}.

A number of studies have been devoted to the electromechanical properties of NBT-$x$BT single crystals, either pure \cite{Ge2008,Ge2009a,Chen2010,Moon2011,Zhang2011} or Mn-doped \cite{Liu2008,Zhang2009,Ge2010,Sun2011}, in the vicinity of the MPB. The highest piezoelectric coefficient obtained so far at ambient conditions has been reported in Mn-doped NBT-10BT poled along $[001]$ and reaches 483~pC/N \cite{Zhang2009}. Mn-doped crystals show a lower depolarization temperature $T_d$ and larger piezoelectric constants than undoped crystals, in contrast with classical lead-based materials where doping with acceptor ions such as Mn$^{2+}$ is known to reduce dielectric losses but usually at the expense of a small decrease of the piezoelectric constants \cite{Priya2002_c}. From a structural point of view, Mn doping has been shown to favor the formation of microsized tetragonal ferroelectric domains \cite{Yao2011}. Here, we study Mn-doped crystals since Mn-doping considerably reduces the dielectric losses and therefore greatly facilitates the poling process.

The aim of this study is to focus on the intrinsinc properties, i.e. the properties determined from a single domain state. Those are essential for a good understanding of the properties of multidomain crystals and quantification of potential enhancement mechanisms. However, since it is commonly assumed (and sometimes experimentally reported) that single domain states are unstable \cite{Ge2008}, these properties have not really been reported yet. Crystals with nominal composition NBT-6\%BT with 0.1\%~Mn were grown at the Shanghai Institute of Ceramics as described in ref. \cite{Ge2008}. The chemical composition was checked by ICP spectroscopy. Various samples were cut along the $[001]$, $[110]$ and $[111]$ directions with dimensions of the order of 4$\times$2$\times$0.5~mm$^2$ (2$\times$1$\times$0.5~mm$^2$ for $[111]$-oriented crystals) and polished with diamond paste. Gold electrodes were sputtered on the relevant faces. Samples were poled in a silicon oil bath, in a procedure very similar to what is usually described in the literature (eg. Refs. \cite{Ge2008,Ge2009,Chen2010}): they were heated up to 130$^\circ$C, then a poling electric field in the range of 30 to 50~kV/cm was applied for 30~min, and the sample was cooled down with the poling field with an approximate speed of 4~K/min. The piezoelectric coefficients $d_{33}$ were measured by the Berlincourt method using a commercial $d_{33}$-meter by Channel Products (load amplitude 250~N, frequency 110~Hz). 

\begin{figure}
\includegraphics[width=0.48\textwidth]{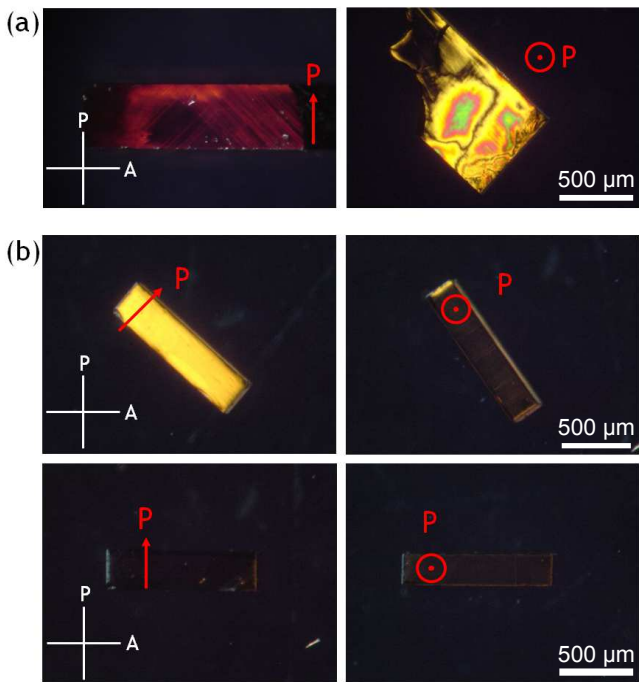}
\caption{Investigation of tetragonal single domain states by optical observation under polarized light. (a) Imperfect single domain obtained after initial poling, showing domain coexistence and interference fringes. (b) Single domain after polishing and repoling at ambient temperature, showing the expected extinctions.}
\label{fig:singledomain}
\end{figure}

Our crystals are on the tetragonal side of the phase diagram, and it was shown in Ref. \cite{Ge2009} that a bias field along $[001]$ stabilizes a tetragonal phase over the rhombohedral variant in Mn-doped NBT-5.6\%BT single crystals. Therefore we can expect to obtain a single domain by poling the crystal along $[001]$. Indeed, we found that $[001]$-poled crystals were close to single domain states, and yet they were imperfect in two ways: i) they were found cracked or broken, probably as a result of the high strain at the ferroelectric transition and ii) the samples were not homogeneous under polarized light; instead the top view showed unexpected interference fringes while the side view reveals the clear presence of thin needle-like domains embedded in an larger homogeneous region (figure \ref{fig:singledomain} (a)). We have performed X-ray diffraction on the $[001]$ face of the sample (cobalt anode with $\lambda = 1.79053$~\AA, penetration depth of less than 1 $\upmu$m). The diffractogram (not shown) clearly revealed the weak but significant presence of a $(400)$ reflection next to the expected $(004)$ reflection, confirming the imperfect poling of the samples. We also estimated the tetragonal lattice constants and the density (see table \ref{tab:properties}). In order to study the properties of the single domain, we therefore processed the samples further by cutting samples in order to isolate nicely poled areas, polishing the surfaces again, and repoling them at ambient temperature with a 30~kV electric field applied during 30 minutes. Final sample volume was approximately one half of the original sample volume. Only after this subsequent treatment could we obtain single domains that nicely show the expected extinctions under polarized light for a tetragonal crystal (figure \ref{fig:singledomain} (b)). We could then proceed to the measurement of the intrinsic properties.

\begin{figure}
\includegraphics[width=0.45\textwidth]{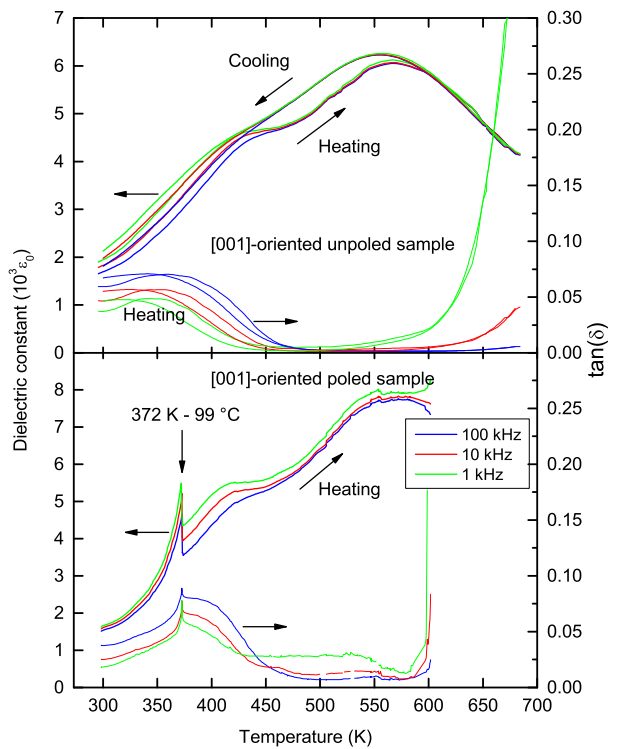}
\caption{Dielectric constant and loss tangent of $[001]$-oriented single crystals as a function of temperature for an unpoled crystal (top) and a poled crystal during depolarization (bottom).}
\label{fig:dielec}
\end{figure}

At room temperature, the dielectric constants $\eps_{33}^T$ and $\eps_{11}^T$ measured at 1~kHz amount to 1520 and 2200 respectively. The temperature dependence of the dielectric constant $\eps_{33}^T$ for a $[001]$-oriented sample is shown in figure \ref{fig:dielec}, both for a poled sample upon heating and for an unpoled sample upon heating and cooling (5~K/min in all cases). For the unpoled sample, the results are fully consistent with previous reports, with a broad maximum at 566~K (heating) and 554~K (cooling) and a relaxation behavior below a second characteristic temperature 467~K. During the depolarization of a poled sample, $\eps_{33}^T$ shows a very sharp anomaly at 372~K (99$^\circ$C). This transition temperature is significantly lower than the 120$^\circ$C reported for pure NBT-BT crystals by Zhang \etal{} using the same technique \cite{Zhang2011}, confirming the tendency of Mn-doping to lower the transition temperature. Let's point out that the depolarization temperature may slightly differ if measured by other techniques \cite{Anton2011}. 

Last, the piezoelectric constant $d_{33}$ measured by the Berlincourt method amounts to 570~pC/N, which is higher than the highest value reported so far. This calls the following comments. i) The special care taken in the preparation of the sample after the initial poling is an important condition of this achievement. Indeed, the $d_{33}$ measured on the sample after the first poling but before the post-treatment was always below 500~pC/N. ii) This is obtained for a single domain crystal, i.e. without any phase coexistence at the macroscopic level, although of course, we cannot exclude the existence of a more complicated structure at the nanoscale below the resolution of our techniques. 

\begin{figure}[h]
\includegraphics[width=0.45\textwidth]{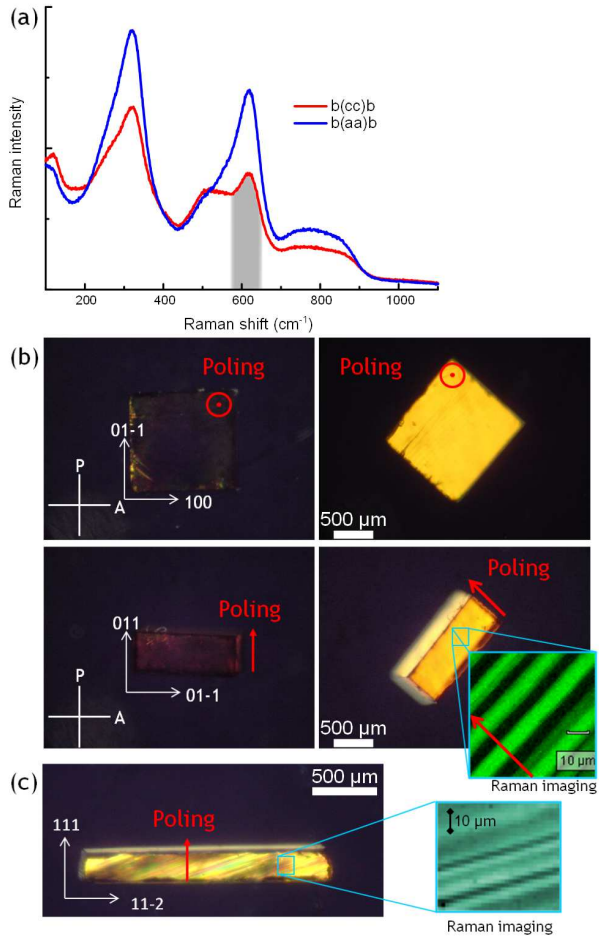}
\caption{(a) Raman spectrum of the single domain state, in the $b(aa)b$ and $b(cc)b$ configurations. (b) and (c) Optical micrographs and Raman mapping of the $[011]$ and $[111]$-poled sample respectively. In both cases, the incident and analyzed lights are polarized vertically.}
\label{fig:multidomain}
\end{figure}

Reference Raman spectra were recorded on the single domain with a Renishaw spectrometer in either blue (488~nm) or green (514~nm) (no wavelength dependence was observed). The spectra, pictured in figure \ref{fig:multidomain} (a), can be described as three broad bands, each divided into two main components. The comparison of the spectra in the $b(aa)b$ and $b(cc)b$ configurations, i.e. with the laser polarization respectively orthogonal and parallel to the polarization direction of the sample, shows the very strong anisotropy of the higher frequency component of each of these bands. This corroborates the analysis by Gregora \etal{} that this high-frequency component arise from a A-symmetry component resulting from the splitting of a polar mode triplet \cite{Gregora2010}. For our purpose, the intensity of this peak can serve as a marker for the imaging of the domain structure of multidomain single crystals. 

In a second step, we investigated samples oriented along a $[011]$ direction, for which a 2T domain structure with two families of tetragonal domains is expected. The samples were poled with the same protocol with a first poling by the field-cooling method and recutting of a clean piece of poled crystal. Repoling at ambient temperature, on the other hand, was found unnecessary. For a tetragonal crystal poled along $[011]$, we may expect a laminar structure of tetragonal domains separated by uncharged (head-to-tail) domain walls perpendicular to the poling direction. Optical microscopy cannot fully confirm this expectation but the extinctions observed under crossed polarizers is consistent with this picture (figure \ref{fig:multidomain} (b)). In addition, the domain structure was imaged by micro-Raman by imaging the intensity of the band at 620~cm$^{-1}$, where the polarization direction of the laser was chosen along a $[001]$ direction. The mapping reveals a clear domain structure with parallel domains stacked along the $[011]$ direction with thickness of the order of 10~$\upmu$m. The piezoelectric constant of this sample amounts to 225~pC/N.

Last, $[111]$-oriented samples were poled with poling fields of 25, 30 and 50~kV/cm. The different poling field did not yield any significant difference on the final state: all exhibited a piezoelectric constant of about 130~pC/N qualitatively similar domain structures under polarized light. The optical inspection shows the three domain families expected for a 3T domain structure with domain wall orientations that are consistent with tetragonal domains separated by uncharged domain walls as described for instance by Erhart and Cao \cite{Erhart2003_b}. In this case, the contrast of Raman spectroscopy for domain structure characterization is much weaker than in the 2T case, since i) the domain structure cannot be considered two-dimensional and ii) the laser does not propagate along a high-symmetry direction of the domains. The map shown in figure \ref{fig:multidomain} nonetheless give evidence for stripes of domain of the order of 10~$\mu$m. We did not observe any significant change in the domain size for the different poling fields used.

\begin{table}
\caption{Summary of the properties of the tetragonal single domain state. Values for the dielectric constants and the loss angles are measured at 1~kHz.}
\label{tab:properties}
\begin{tabular}{l l l l l}
\hline\hline\noalign{\smallskip}
$a$ 													& 3.892 & \AA \\
$c$ 													& 3.940 & \AA \\
$\rho$ 												& 5.93 	& g.cm$^{-3}$ \\
$\eps_{11}^T$				 					& 2200 	& $\eps_0$ & $\tan \delta_{11}$ 						& 0.010 \\
$\eps_{33}^T$				 					& 1520 	& $\eps_0$ & $\tan \delta_{33}$ 						& 0.018 \\
$d_{33}$ 											& 570 	& pC/N \\
\noalign{\smallskip}\hline\hline
\end{tabular}
\end{table}

In a domain-engineering perspective, let us recall that the effective piezoelectric properties of a domain-engineered crystal combines an intrinsic contribution that can be simply calculated from the intrinsic properties of the single domain by rotation of the piezoelectric tensor, and extrinsic contributions relative to the domain structure, presence of domain walls etc. Ideally, an optimized material should combine both phenomena. For the crystals studied here, the dielectric anisotropy, quantified by the ratio $\eps_{11}^T/\eps_{33}^T=1.45$, is low and classifies this crystal in the family of so-called "extender piezoelectrics" \cite{Davis2007a}. Such piezoelectrics are characterized by a relatively low piezoelectric constant $d_{15}$ and exhibit their higher piezoelectric response along the poling direction, as is actually seen in our study. Interestingly, the dielectric anisotropy of PZN-12PT is much higher than for PbTiO$_3$, whereas it is the opposite with NBT-6BT as compared to BaTiO$_3$, which is again an indication of the very different mechanisms at work in the two systems. As far as the extrinsic contribution is concerned, effects have been experimentally demonstrated in BaTiO$_3$ poled along $[111]$ by Wada and coworkers \cite{Wada2004,Wada2005_c}, and have been a topic of various calculations and simulations \cite{Hlinka2006,Rao2007,Li2012}. Yet, for the crystals considered here, the observed domain sizes could not be significantly altered by the different poling process used, and remain well above the limit where a domain-size enhancement would be expected from the simulations.

In summary, we have successfully prepared Mn-doped piezoelectric NBT-6\%BT single crystal in the 1T, 2T and 3T configurations. We have shown that a single domain state 1T can be stabilized if special care is taken during sample preparation. In this single domain state, the piezoelectric constant reaches 570~pC/N, the highest value measured so far in this family.  Piezoelectric properties of the domain-engineered single crystals on the other hand show inferior $d_{33}$, which reflects the extender character of those tetragonal crystals. This implies that little improvement is to be expected from domain engineering in classical sense, in contrast to tetragonal crystals of the lead-based families, unless more elaborate poling protocols are explored.

The authors are thankful to Ji\v r\'i Erhart and Stanislav Pano\v s from the Technical University of Liberec for preliminary poling tests, as well as J. Kope\v cek, K. Jurek and I. Gregora for support and complementary analyses. Financial support of the Grant Agency of the Czech Republic (project No. P204/10/0616) and from the MPO project FR-TI2/165 is acknowledged. 

{\it Note added to the proof:} A similar study was independently carried out at SIC CAS. Occasionally, the $d_{33}$ value of NBT-6\%BT recorded by Berlincourt meter was even exceeding the 570~pC/N reported here.

\end{document}